\documentclass[12pt]{article}
\usepackage{amsmath}
\usepackage{epsfig}
\usepackage{amstext,amssymb,amsfonts}

\advance\voffset by -2.0cm
\advance\hoffset by -1.75cm
\def\Tr{\,{\rm Tr}\, }

\def\be{\begin{equation}}
\def\ee{\end{equation}}
\def\ba{\begin{eqnarray}}
\def\ea{\end{eqnarray}}

\renewcommand{\H}{{\cal H}}

\newcommand{\M}{{\cal M}}
\newcommand{\N}{{\cal N}}

\newcommand{\eg}{{\it e.g.~}}
\newcommand{\ie}{{\it i.e.~}}

\newcommand{\Nop}{\mathbb{N}}

\begin{document}
\vspace*{1.5cm}

\begin{center}
{\Large 
{\bf Modular differential equations and null vectors}}
\vspace{2.5cm}

{\large Matthias R.\ Gaberdiel$^{1}$}
\footnotetext{$^{2}${\tt E-mail: gaberdiel@itp.phys.ethz.ch}} 
{\large and} 
{\large Christoph A.\ Keller$^{2}$}
\footnotetext{$^{3}${\tt E-mail: kellerc@itp.phys.ethz.ch}} 
\vspace*{0.5cm}

Institut f{\"u}r Theoretische Physik, ETH Z{\"u}rich\\
CH-8093 Z{\"u}rich, Switzerland\\
\vspace*{3cm}

{\bf Abstract}
\end{center}
We show that every modular differential equation of a rational conformal 
field theory comes from a null vector in the vacuum Verma module. We
also comment on the implications of this result for the consistency
of the extremal self-dual conformal field theories at $c=24 k$.


\newpage
\renewcommand{\theequation}{\arabic{section}.\arabic{equation}}


\section{Introduction}
\setcounter{equation}{0}

Every rational conformal field theory possesses a modular differential equation.
This is to say, the different characters of the finitely many irreducible
highest weight representations satisfy a common differential equation
in the modular parameter. This fact was first observed, using
the transformation properties of  the characters under
the modular group, in \cite{Eguchi:1986sb,Anderson:1987ge,MMS,MMS1}; later
developments of these ideas are described in 
\cite{Eholzer,ES,Flohr:2005cm,Bantay1}.
Following the work of Zhu \cite{Zhu}, the modular transformation properties
of the characters were derived from first principles (see also 
\cite{Nahm:1991ie}). Zhu's derivation suggests that the modular
differential equation is a consequence of a null-vector relation in 
the vacuum Verma module \cite{Gaberdiel:2007ve}, see also
\cite{Flohr:2005cm}. In this paper we shall show that
this idea is indeed correct.
\medskip

The recent interest in this problem arose from the analysis of
Witten concerning pure gravity in AdS$_3$ \cite{Witten:2007kt}. 
He suggested that the corresponding boundary theories should
be holomorphically factorising bosonic conformal field theories at 
$c=24 k$ with $k=1,2,\ldots$, where $k\rightarrow \infty$ describes
the classical limit of the AdS$_3$ theory. Furthermore, the
corresponding chiral theories should be extremal, meaning that up to
level $k+1$ above the vacuum, the theory only consists of Virasoro  
descendants of the vacuum state. For $k=1$, the resulting 
conformal field theory is the famous Monster theory
\cite{FLM,Borcherds} (for a beautiful introduction see \cite{Gannon}),
but for $k\geq 2$ an explicit realisation of these theories is so far 
not known. The above constraints, however, specify the character
of these meromorphic conformal field theories uniquely 
\cite{Hoehn,Witten:2007kt}. Furthermore,  the $k=2$ vacuum
amplitudes are well-defined on higher genus Riemann surfaces
\cite{Witten:2007kt}, and the genus $2$ amplitude of the
$k=3$ theory was shown to be consistent (by some other methods)
\cite{Gaiotto:2007xh}. Their method determines also the genus 2
partition functions uniquely up to $k\leq 10$.  There has also been
some evidence that suggests that these extremal theories account
correctly for the corresponding gravity amplitudes
\cite{Manschot:2007zb,Yin:2007gv,Yin:2007at,Maloney:2007ud}.
\smallskip

On the other hand, it is not clear whether the theories with $k\geq 2$
do indeed exist. It was proposed  in \cite{Gaberdiel:2007ve}
that the analysis of their modular differential equations could shed light
on this question. Since these theories are self-dual, they
only have a single highest weight representation, and thus only a 
single character. One can then obtain an estimate for the order $s$
of the differential equation that annihilates this character; this 
is proportional, for large $k$, to $s\sim \sqrt{k}$.  On the other hand, 
if there is a direct relation between modular differential equations
and null-vectors in the vacuum Verma module, a modular differential
equation at order $s$ should imply that
the vacuum Verma module has a null-vector at conformal weight $2s$.
This would then lead to a contradiction for $k\geq 42$ 
since the extremal theories do not have any null-vectors at such low
levels \cite{Gaberdiel:2007ve}. 

In \cite{Gaberdiel:2007ve} a more specific conjecture was made
(and supported by some evidence). It was suggested that 
if a conformal field theory satisfies an order $s$ modular differential
equation, then $L_{-2}^s\Omega\in O_{[2]}$. (In particular, this 
conjecture implies the weaker statement that the vacuum Verma 
module has a  null-vector at level $2s$.) The original form of the 
conjecture has turned out to be incorrect: the example of 
Gaiotto \cite{Gaiotto:2008jt} involving tensor products of the Monster
theory demonstrates this fact. This example is however
not in conflict with the weaker statement that the vacuum 
Verma module possesses a null-vector at  level $2s$ --- albeit
one that is of a somewhat different form. In fact, the tensor 
products of the Monster theory have many null-vectors at levels
that are even below the one suggested by the order of the modular 
differential equation!
\smallskip

In this paper we want to analyse the relation between
modular differential equations and null-vectors in detail. 
One of our main results is that every modular differential 
equation comes from a null-vector in the vacuum Verma module 
(see (\ref{nullv})). We shall explain under which conditions
this leads to a relation of the form $L_{-2}^s\Omega\in O_{[2]}$,
thus giving in particular a null-vector at level $2s$. We shall
also explain in detail how the counterexample of Gaiotto
avoids this conclusion; as we shall see, this is intimately related
to the fact that the tensor product of two (or more) Monster theories 
has many other null-vectors. We also comment on the fact that the
existence of these additional null-vectors can be seen
from an analysis of the Monster theory character; 
the same is true for Witten's theory at $k=2$, but, at least from
the point of view of the character, there are no indications that
the theories with $k\geq 3$ should have sufficiently many
null vectors to avoid a contradiction along these lines.
\bigskip

The paper is organised as follows. In section~2, we explain in a 
comprehensive manner how the modular differential equation arises from
the analysis of Zhu \cite{Zhu}. We also illustrate this with a simple and
very explicit example, the Yang-Lee model at $c=-22/5$. In section~3
we show that a modular differential equation always leads to 
a null-vector relation in the vacuum representation. We analyse under
which conditions this implies that $L_{-2}^s\Omega\in O_{[2]}$, and
how the counterexample of Gaiotto avoids this conclusion. 
Finally, we comment in section~4 on the implications of these
considerations for the existence of the extremal self-dual conformal
field theories at $k\geq 42$. 
In order to be comprehensive we have included a number of
appendices: in appendix~A we give a brief introduction to Zhu's
algebra, while the torus recursion relations that underly the torus
analysis of Zhu are derived (in a physicists manner) in
appendix~B. Appendix~C  describes our conventions for the
Weierstrass functions and Eisenstein series, and details their modular
properties, while appendix~D describes one of the technical arguments
of the paper.

\section{The modular differential equation}
\setcounter{equation}{0}

Let us begin by explaining the structure of torus amplitudes in a 
rational conformal 
field theory. It is usually believed (and it follows in fact from the
analysis of Zhu   
\cite{Zhu}) that the torus amplitudes can be described in terms of the
characters of the highest weight representations of the conformal
field theory. These characters satisfy a modular differential equation
\cite{MMS,MMS1} (for earlier work see
\cite{Eguchi:1986sb,Anderson:1987ge}). In this section we want to
explain the origin of this differential equation from the point of
view of Zhu \cite{Zhu}. 

Let $V$ be a meromorphic conformal field theory (or vertex operator
algebra). For each state $a\in V$ we have a vertex operator $V(a,z)$,
whose modes we denote by $a_n$ (using the usual physicists'
conventions). The zero mode of $a$ plays a special role, and we denote
it by $o(a)\equiv a_0$. On the torus, it is more advantageous
to use different coordinates; the associated modes are then denoted by
$a_{[n]}$. All of this is explained in more detail in
appendix~\ref{app:zhu}.

It follows from an elementary (but somewhat tedious) calculation due to 
Zhu (Proposition 4.3.5 of \cite{Zhu} --- we sketch an outline of the
argument in appendix~\ref{app:torus}) that 
\be\label{rec1} 
{}\Tr_\H \Bigl(o(a_{[-h_a]}b)\, q^{L_0} \Bigr) = 
\Tr_\H \Bigl(o(a)\, o(b) \, q^{L_0} \Bigr)
+ \sum_{k=1}^\infty G_{2k}(q) 
\Tr_\H \Bigl( o(a_{[2k-h_a]}b) \, q^{L_0} \Bigr)\ . 
\ee
Here the trace is taken in any highest weight representation $\H$ of the
chiral algebra, and $G_{n}(q)$ denotes the $n^{\rm th}$ Eisenstein
series; our conventions for the  Eisenstein series (as well as their
main properties) are summarised in appendix~\ref{app:Eisenstein}. 
Next we apply (\ref{rec1}) with $a$ replaced by $L_{[-1]}a$, and 
use that $(L_{[-1]}a)_{[n]}= -(h_a+n)a_{[n]}$ (as follows from
(\ref{A4}) upon taking a derivative), as well as 
$o(L_{[-1]}a) = (2\pi i)\, o(L_{-1}a + L_0 a) = 0$, which is a
consequence of (\ref{A14}); this leads to 
(see Proposition 4.3.6 of \cite{Zhu}) 
\be\label{rec2} 
{}\Tr_\H \Bigl(o(a_{[-h_a-1]}b) \, q^{L_0}\Bigr) 
+ \sum_{k\geq 1} (2k-1)
G_{2k}(q)\, \Tr_\H \Bigl( o(a_{[2k-h_a-1]}b) \, q^{L_0}\Bigr) = 0\ . 
\ee
The term with $k=1$ does not contribute here since the 
trace of $o(a_{[-h_a+1]}b)$ vanishes --- as follows from
(\ref{commutatora}) in the appendix, it is a commutator and hence
vanishes in the trace.

Equation (\ref{rec2}) motivates now the
following definition. Let $V[G_4(q),G_6(q)] $ be the space 
of polynomials in the Eisenstein series with coefficients in $V$. 
This is a module over the ring $R=\mathbb{C}[G_4(q),G_6(q)]$ which
carries a natural grading given by the modular weight of each
monomial; since $G_4$ and $G_6$ generate all modular forms, we have in
particular that $G_{2k}(q)\in R$ for $k\geq 2$.
Then we define $O_q(V)$ to be the submodule of $V[G_4(q),G_6(q)]$ 
generated by states of the form
\be\label{oqdef}
O_q(V): \qquad
a_{[-h_a-1]}b + \sum_{k\geq 2} (2k-1)\, G_{2k}(q)\,
a_{[2k-h_a-1]}\,b\ ,
\ee
where $a,b\in V$. Here the sum is finite, as $a_{[n]}$ annihilates $b$
for sufficiently large $n$. By (\ref{rec2}), it is now clear that  
\be\label{oqp}
\Tr_\H\Bigl( o(v)\, q^{L_0}\Bigr)=0  \qquad \hbox{if $v\in O_q(V)$.}
\ee
This is true for every character, \ie independent of the highest weight 
representation $\H$ that is being considered. 
For later convenience we also note that
\be
a_{[-h_a-n]}b  
- (-1)^n \sum_{2k\geq n+1} \binom{2k-1}{n}
G_{2k}(q)\, a_{[2k-h_a-n]}b \in O_q(V) \ , \quad
\forall n\geq 1\ , \label{rec3}  
\ee
as can be seen by evaluating the above identity repeatedly
with $a$ being replaced by $L_{[-1]}a$.
\medskip
 
Suppose now that for a conformal field theory we can find an integer
$s$ and modular forms $g_r(q)$ of weight $2(s-r)$ such that
\footnote{As is explained   
in \cite{Zhu}, the existence of such a vector follows for example from
the $C_2$ condition that is  
believed to hold for every rational conformal field theory --- see
also appendix~A.2.}
\be
\bigl(L_{[-2]}\bigr)^s \Omega +
 \sum_{r=0}^{s-2} g_r(q) \,\bigl(L_{[-2]}\bigr)^r \Omega \in
O_q(V)\ . \label{oqstate}
\ee
We then claim that all the 
characters $\chi_\H(q) := \Tr_\H\bigl(q^{L_0 -\frac{c}{24}} \bigr)$ of
the conformal field theory satisfy a common modular 
covariant 
differential equation, \ie an equation of the form
\be\label{mde}
\left[ D^s + \sum_{r=0}^{s-2} f_r(q) D^r\right]\chi_M(q) = 0\ .
\ee
Here $D^s$ is the order $s$ differential operator (see
appendix~\ref{app:Eisenstein})
\be
D^s = D_{2s-2} \, D_{2s-4} \cdots D_2\, D_0 \ , \qquad \hbox{with} \qquad
D_r = q \frac{d}{dq} - \frac{r}{4\pi^2} G_2(q)
= q \frac{d}{dq} - \frac{r}{12} E_2(q)  \ , 
\ee
and $f_r(q)$ is a modular form of weight $2(s-r)$.\footnote{We shall
use two different conventions for the Eisenstein series, namely
$G_n(q)$ and $E_n(q)$, in this paper; the two functions only differ by
an overall normalisation constant, see appendix~C.}

To show this, note that because of the defining property of $O_q(V)$ 
(\ref{oqp}), we know that the character of the zero mode of the left
hand side of (\ref{oqstate}) vanishes. On the other hand, each term in
this expression can be turned into a differential operator 
\be\label{dop}
\Tr_\H\Bigl( o\bigl((L_{[-2]})^r\Omega\bigr)\, q^{L_0-\frac{c}{24}}\Bigr) 
= P_r(D) \Tr_\H \Bigl(q^{L_0-\frac{c}{24}} \Bigr) \ ,
\ee
where $P_r(D)$ is a modular covariant differential operator of order
$r$ with modular weight $2r$. To see
(\ref{dop}) we note that for $r=1$ we obtain directly
\be
\Tr_\H \Bigl(o(L_{[-2]}\Omega) \, q^{L_0-\frac{c}{24}} \Bigr)
= (2\pi i)^2 
\Tr_\H \Bigl(\bigl(L_0-\frac{c}{24} \bigr)\, q^{L_0-\frac{c}{24}} \Bigr) 
= (2\pi i)^2 \Bigl(q\frac{d}{dq}\Bigr)  
\Tr_\H \Bigl( q^{L_0-\frac{c}{24}} \Bigr) \ ,
\ee
which is modular covariant since the character has modular weight 0.
The case of general $r$ follows by applying (\ref{rec1}) (which
clearly still works if we replace $q^{L_0}$ by $q^{L_0-c/24}$)  
\begin{eqnarray}
{\displaystyle
\Tr_\H\Bigl(o(L_{[-2]} (L_{[-2]})^r \Omega) \, q^{L_0-\frac{c}{24}} \Bigr)}
& = & {\displaystyle
(2\pi i)^2   q \frac{d}{dq} 
\Tr_\H \Bigl( o( (L_{[-2]})^r \Omega)\, q^{L_0-\frac{c}{24}} \Bigr)} 
\nonumber \\
& & \ 
+ 2 r G_2(q) \, \Tr_\H \Bigl( o( (L_{[-2]})^r \Omega)\, 
q^{L_0-\frac{c}{24}} \Bigr) 
  \\
 & & \
 + \sum_{k\geq 2} G_{2k}(q)\, 
 \Tr_\H \Bigl( o( L_{[2k-2]}(L_{[-2]})^r \Omega)\, 
q^{L_0-\frac{c}{24}} \Bigr) \ . 
 \nonumber
\end{eqnarray}
In the last line we commute the positive $L_{[2k-2]}$ modes to the
right, using the Virasoro commutation relations. The final result is a
vector of the form $(L_{[-2]})^{r+1-k}\Omega$, which leads to a
differential operator of lower order, multiplied by the modular form
of appropriate weight. The first two terms, on the other hand,
just produce the  covariant derivative $D_{2r}$ for a form of weight
$2r$. Collecting all terms, we get the desired operator $P_r(D)$. Note
that the leading term of $P_r(D)$ is proportional to $D^r$; for
the first few values of $r$, the explicit formula for $P_r(D)$ is
given in appendix~B.1. This completes the derivation of the 
modular differential equation.

\subsection{A simple example}

Let us illustrate this construction with a simple example, the 
Yang-Lee minimal model at $c=-\frac{22}{5}$. This is the `simplest'
minimal model since it only has two highest weight representations,
the vacuum representation at $h=0$ as well as the representation at
$h=-\frac{1}{5}$. The vacuum representation has a null-vector 
at level $4$,
\be\label{YLnull}
{\cal N} = \left( L_{[-4]}-\frac{5}{3}L^2_{[-2]}\right) \Omega\  .
\ee
We want to use $\N$ to obtain an expression of the form
(\ref{oqstate}). To this end we observe that 
$L_{[-4]}\Omega$ is already in $O_q(V)$ since (\ref{rec3}) implies
that 
\be
O_q(V) \ni L_{[-4]}\Omega - \sum_{k\geq 2} \binom{2k-1}{2}
G_{2k}(q)\, L_{[2k-4]}\Omega = L_{[-4]}\Omega\ .
\ee
Since $\N$ is a null-vector, the sought-after relation is then simply 
\be
L_{[-2]}^2 \Omega \in O_q(V)\ .
\ee
Using the explicit expression for (\ref{dop}) derived
in appendix~\ref{app:diffops}, we obtain the differential
equation
\be\label{YLmod}
0 = \Tr_\H\Bigl( o(L_{[-2]} L_{[-2]}\Omega)\, 
q^{L_0-\frac{c}{24}} \Bigr)
=(2\pi i)^4 \left[ D^2 
 - \frac{11}{3600}\,  E_4(q)\right]\, \chi_\H(q)\ .
\ee 
The two characters of the Yang-Lee model are explicitly given as 
(see for example \cite{DiF})
\begin{eqnarray}
\chi_0(q) & = & \frac{1}{\eta(q)} \sum_{n\in {\mathbb Z}} 
\left( q^{\frac{(20n-3)^2}{40}} - q^{\frac{(20n+7)^2}{40}} \right)
\\
\chi_{-1/5}(q) & = & \frac{1}{\eta(q)} \sum_{n\in {\mathbb Z}} 
\left( q^{\frac{(20n-1)^2}{40}} - q^{\frac{(20n+9)^2}{40}} \right) \ , 
\end{eqnarray}
where $\eta(q)$ is the usual Dedekind eta function
\be
\eta(q) = q^{\frac{1}{24}} \prod_{n=1}^{\infty} (1-q^n) \ . 
\ee
One easily checks (using for example Mathematica) that the two
characters are indeed the two solutions of this second order
differential equation. We have also performed the analogeous analysis
for the Ising model.

\subsection{Relation to the null-vector}

In the above example, the vector of the form (\ref{oqstate}) in $O_q(V)$
was a direct consequence of a null-vector relation in the vacuum
representation, see (\ref{YLnull}).  This is actually generally true: a 
vector of the form (\ref{oqstate}) in $O_q(V)$ can only exist if the 
vacuum representation has a null-vector at level $2s$. To see this 
we recall that $V[G_4(q),G_6(q)]$ carries two
grades: the {\it conformal weight} of the vectors in $V$ (with respect
to $L_{[0]}$), and the {\it modular weight} of the coefficient
functions (that are polynomials in $G_4$ and $G_6$). Furthermore, the
relations that define $O_q(V)$ are homogeneous with respect to the
grade that is the sum of these two grades, as is manifest from
(\ref{oqdef}).  

Since the relation (\ref{oqstate}) is a relation in $V[G_4(q),G_6(q)]$ it
must hold separately for every conformal weight and every modular
weight. If we consider the component at conformal weight $2s$ and
modular weight zero, we therefore get a relation of the form
\be\label{null}
\bigl( L_{[-2]} \bigr)^s \Omega + \sum_j a^{j}_{[-h(a^j)-1]} \, b^{j} = 0  \ ,
\ee
where $h(a^j)$ is the conformal weight (with respect to $L_{[0]}$) of
$a^{j}$.  This is necessarily a non-trivial relation in the Verma
module since $L_{[-2]}$ is not of the form $a_{[-h_a-1]}$ for any
$a$. Such a non-trivial relation is usually called a null-vector. We
mention in passing that it implies that 
$\bigl(L_{[-2]} \bigr)^s \Omega$ vanishes in the $C_2$ quotient space
of Zhu (that is briefly discussed in appendix~A.2), as was already
mentioned in \cite{Gaberdiel:2007ve}.

\section{Reconstructing the null-vector}
\setcounter{equation}{0}

As we have seen above, a vector of 
the form (\ref{oqstate}) in $O_q(V)$ implies that the characters
of the theory satisfy a common order $s$ modular differential equation.
We have also shown that such a relation in $O_q(V)$ can only 
exist if the vacuum representation has a null-vector at conformal weight
$2s$, see (\ref{null}).

We would now like to show a partial converse to these statements,
namely that every modular differential equation implies that the 
vacuum Verma module has a null-vector. We shall assume that 
Zhu's algebra is semisimple,
as is known to be the case for rational conformal field theories (in
the mathematical sense) \cite{Zhu}. In particular, this is the case
for the self-dual theories, for which Zhu's algebra is
one-dimensional, consisting only of the identity. In this section we
only sketch the idea of the proof; more details of the calculation
can be found in appendix~\ref{app:convergence} .

\subsection{The underlying vector}

Suppose now that we have a modular covariant differential equation of 
the form (\ref{mde}) that annihilates all characters of the conformal field
theory. Using the arguments of section~2 in reverse order, it is easy to 
see that there is then a vector $K(q)$ of the form 
\be\label{kdef}
K(q) \equiv \bigl(L_{[-2]}\bigr)^s \Omega +
 \sum_{r=0}^{s-2} g_r(q)\, \bigl(L_{[-2]}\bigr)^r \Omega 
\ee
where each $g_r(q)$ is a modular form of weight $2(s-r)$, 
that has the property that 
\be\label{relation}
\Tr_\H \Bigl( o(K(q)) \, q^{L_0-\frac{c}{24}} \Bigr)=0
\ee
for all characters of the conformal field theory. 
Let us consider the limit
\be
\lim_{q\rightarrow 0} q^{\frac{c}{24}-h}  \Tr_\H \Bigl( o(K(q)) \, 
q^{L_0-\frac{c}{24}} \Bigr)=0 \ , 
\ee
where $h$ is the conformal weight of the highest weight
state in $\H$. In this limit only the highest weight states $\H^0$ 
in $\H$ contribute, and we conclude that 
\be
\Tr_{\H^{0}} \Bigl( o(K(0)) \Bigr) = 0 \ . 
\ee

\subsection{Using Zhu's Theorem} \label{ss:ZhuTheorem}

The above argument has shown that $K(0)$ acts trivially in the trace
of an arbitrary highest weight representation. The action of the
elements of $V$ on highest weight states is captured by Zhu's algebra
(for a brief introduction see appendix~\ref{app:zhu}.1). If Zhu's
algebra is semisimple (as we shall assume) then the fact that $K(0)$
is trivial in all traces implies that $K(0)$ must equal a commutator
in Zhu's algebra; this is shown in appendix~\ref{app:convergence}. 
This implies that up to commutator terms of the form
$d^l_{[-h(d^l)+1]}\, e^l$,  
$K(0)$ lies in $O_{[1,1]}$, the subspace by which we quotient to
obtain Zhu's algebra $A(V)$. On the other hand, $O_{[1,1]}$ is closely
related to $O_q(V)$, since again
up to a commutator terms, every element in $O_{[1,1]}$ can be obtained
as the limit $q\rightarrow 0$ of an element $H_j(q) \in O_q(V)$ (see
again appendix~D).

\noindent Taking these statements together they now imply that 
$K(0)$ can be written as
\be
K(0)-\sum_l d^l_{[-h(d^l)+1]}\, e^l-\sum_j H^j(0) = 0\  ,
\ee
where each $H^j(q)$ is an element of $O_{q}(V)$
\be
H^j(q) = a^j_{[-h(a^j)-1]}b^j + \sum_{k\geq 2} (2k-1)G_{2k}(q)\,
a^j_{[2k-h(a^j)-1]}b^j \in O_q(V)\  \label{hdef}
\ee
for some suitable set of $a^j$ and $b^j$. Next we define an element
$N(q) \in V[G_4,G_6]$ by  
\be\label{ndef}
N(q)\equiv K(q)- \sum_l d^l_{[-h(d^l)+1]}\, e^l - \sum_j H^j(q)\ .
\ee
By construction, $N(0)=0$, and hence $N(q)$ is proportional to $q$. We
can then divide by $q$, and repeat the above argument. Recursively
this allows us to prove that 
\be\label{ndef1}
K(q) = \sum_l f_l(q) \, d^l_{[-h(d^l)+1]}\, e^l 
+ \sum_j h_j(q) \, H^j(q) 
\ee
in $V[q]$, where $V[q]$ consists of vectors in $V$ with coefficients
that are formal power series in $q$. If we assume that the theory is
$C_2$-finite (as is expected to be the case for any rational theory)
one can show that only finitely many terms appear and that the    
power series have a non-trivial radius of convergence; this is
explained in appendix~\ref{app:convergence}.
\smallskip

Putting everything together, we can now use (\ref{ndef1}) as well as 
(\ref{kdef}) and (\ref{hdef}) to arrive at the identity 
\begin{eqnarray}\label{nullv}
& & (L_{[-2]})^s \Omega + \sum_{i=0}^{s-1} g_i(q)(L_{[-2]})^i \Omega
\label{L-2} \\
& & \quad = \sum_l f_l(q) d^l_{[-h(d^l)+1]}e^l 
+\sum_{j} h_j(q) \Bigl(a^j_{[-h(a^j)-1]}b^j + \sum_{k\geq 2}
(2k-1)G_{2k}(q)\, a^j_{[2k-h(a^j)-1]}b^j\Bigr)\  \nonumber
\end{eqnarray}
as a relation in $V[q]$. This defines 
the sought after `null-vector' relation in the vacuum Verma
module. Obviously, the full expression is not homogeneous with respect to
conformal weight, and therefore each component ({\it i.e.}\ the terms
of each fixed conformal weight) must vanish separately (and indeed for 
any power of $q$). Some of these relations may be trivial in the Verma 
module, but not all of them can if the original 
modular differential equation from which we started was non-trivial.

\subsection{Consequences}

We have thus shown that every modular differential equation comes from
a null-vector in the vacuum Verma module. We would now like to obtain
more detailed information from (\ref{nullv}). For the application to
the extremal self-dual conformal field theories, it is for instance
also important to determine the conformal weights of the constituent
null-vectors. In particular, one may expect that the term of highest
conformal weight on the left-hand-side --- this is the vector
$(L_{-2})^s\Omega$ --- should be part of a non-trivial null-vector
relation. 

In order to motivate this proposal we observe that the 
coefficients of the vectors of the left-hand-side of (\ref{nullv}) are
all analytic functions in $q$ on the unit disc, $|q|<1$. Therefore the
same has to be true for the coefficients on the
right-hand-side. Generically, one should then expect that the functions
$f_l(q)$ and $h_j(q)$ will also be analytic functions on $|q|<1$; as
we shall discuss later on, there are however situations where this is
not the case.

Now we recall that $V[G_4(q),G_6(q)]$ has two gradings, namely
the ones given by conformal weight and modular weight. By construction 
$(L_{[-2]})^s \Omega$ has modular weight $0$ and conformal weight
$2s$. If $f_l(q)$ and $h_j(q)$ are indeed analytic, then the only
terms of modular weight 0 on the right hand side of (\ref{L-2}) have
constant coefficients.  Moreover, comparing the conformal weights,
only terms of $L_{[0]}$-weight $2s$ can contribute. Thus we can
conclude that we have an identity of the form
\be
(L_{[-2]})^s \Omega = \sum'_{j}a^j_{[-h(a^j)-1]}b^j +
\sum'_{l} d^l_{[-h(d^l)+1]}e^l \ , \label{main2}
\ee
where the prime over the sum indicates that we only include states of 
$L_{[0]}$-weight $2s$, \ie terms with $h(a^j)+h(b^j)+1=2s$ and
$h(d^l)+h(e^l)-1=2s$. Because of the `commutator terms', \ie
the first sum in (\ref{main2}), this identity does not quite imply that
$L_{[-2]}^s\Omega\in O_{[2]}$. However, for the case of the
extremal self-dual theories at $c=24 k$ we can show (see
section~4 below) that this is so, and hence that (\ref{main2}) defines 
indeed a non-trivial null-vector relation. 
\smallskip

In the above argument we have used that there are no holomorphic
functions of negative modular weight; in particular, this implied that 
$h_j(q) G_{2k}(q)$ had modular weight greater or equal to $2k$, and
hence could not contribute to the identity (\ref{main2}). However, as soon
as we allow $h_j$ to be meromorphic, we can no longer guarantee
this. For example, we can then construct other contributions to
(\ref{main2}) from terms with $k\neq 0$ by choosing 
$h_j(q)= G_{2k}(q)^{-1}$. We will now discuss an example of such
a situation.

\subsection{A counterexample}

It was observed in \cite{Gaiotto:2008jt} that for the tensor product
of two (or more) Monster theories, there exist modular differential
equations that do not come from relations of the type
(\ref{main2}). As we shall explain in the following, this
`counterexample' to (\ref{main2}) can be traced back to the failure of
$h_j$ to be holomorphic. We shall also see that this is only
compatible with the holomorphicity of (\ref{nullv}) because the
Monster theory (and indeed the tensor products of the Monster theory)
has many other null-vectors at low levels. These null-vectors are
necessary to guarantee that the apparent non-holomorphic terms on the
right-hand-side of (\ref{nullv}) in fact vanish in the vacuum
representation. Thus it seems that (\ref{main2}) can only be avoided
if the theory has other non-trivial null-vectors at low levels.

\subsubsection{The Monster theory}

To set up the notation we first recall a few facts about the 
case of a single Monster theory; for an introduction to these
matters see for example \cite{Gannon}.  The 
Monster theory has no fields of conformal dimension one, and 
$196884$ fields of conformal dimension $2$. The latter consist of the 
stress-energy tensor whose modes $L_n$ satisfy a Virasoro algebra at
central charge $c=24$,  
\be
[L_m,L_n] = (m-n) L_{m+n} + 2 m (m^2-1) \delta_{m,-n} \ .
\ee
The remaining $196883$ fields $W^i$ transform in an irreducible
representation of the Monster group and satisfy the commutation
relations
\begin{eqnarray}
{}[L_m,W^i_n] & = & (m-n) W^i_{m+n} \nonumber \\
{}[W^i_m,W^j_n] & = & \frac{1}{6} \delta^{ij} m (m^2-1) \delta_{m,-n} 
+ \frac{1}{12} \delta^{ij} (m-n) L_{m+n} \nonumber \\
& & \qquad
+ h^{ij}_{k} (m-n) W^k_{m+n} + f^{ij}_{\alpha} V^\alpha_{m+n} \ , 
\end{eqnarray}
where $V^\alpha_l$ are the modes of the primary fields at conformal
weight three that transform in the $21 296 876$-dimensional  
irreducible representation of the Monster group. The coefficients
$h^{ij}_{k}$ are totally symmetric in all three indices, and define
the structure constants of the so-called Griess algebra. In our
conventions, the metric on the space of the $W^i$ fields is
orthonormal, so we can raise and lower the $i,j,k$ indices freely.

The Monster conformal field theory has many non-trivial relations; the first
non-trivial relation already occurs at level four since we have the
identity (see for example \cite{M})\footnote{This follows from the
equation after (2.9) in \cite{M} upon rewriting his modes $x^i$ with
$i=1,\ldots,196884$ in terms of the $W^i$ and $L$.}
\be \label{n4}
{\cal N}_4 = L_{-2}^2 \Omega + \frac{36}{11} L_{-4}\Omega - 
\frac{12}{30503} \sum_i W^i_{-2} W^i_{-2} \Omega = 0 \ .
\ee
This null-relation does, however, not directly lead to a differential
equation since it is not of the form (\ref{main2}). As was
already explained in \cite{Gaberdiel:2007ve}, the character of the
Monster theory $\chi_M(q)$ satisfies only a third order differential
equation  
\be\label{Monstd}
\left[D^3 + \frac{16}{31} E_6(q) 
- \frac{290}{279} E_4(q) \, D \right] \chi_{M}(q) = 0 \ . 
\ee 
This differential equation can be obtained from the null-vector at
level six (see again \cite{M})
\be\label{N6}
{\cal N}_6 = 
L_{-2}^3 \Omega + \frac{41}{8} L_{-3}^2\Omega + \frac{15623}{1488} L_{-4}
L_{-2} \Omega + \frac{873}{31} L_{-6} \Omega 
- \frac{1}{124} \sum_i W^i_{-4} W^i_{-2}\Omega = 0 \ . 
\ee
In fact, it is easy to see that evaluating the trace of 
$V_0({\cal  N}_6)$ as in section~2 (where in the definition of 
${\cal N}_6$ we replace the $L_{-n}$ modes by $L_{[-n]}$ modes,
and similarly for the $W^i_{-n}$)
leads to the above modular differential equation. (In order to do this
calculation, one also needs to use the commutation relations of the
$W^i$-modes.) 
\medskip

\noindent
There is an independent null-vector at level eight, which is of the
form\footnote{Such a null-vector must exist since, 
up to level $10$, all states that are Monster
invariant can be expressed in terms of Virasoro descendants of the
vacuum. The coefficients can then be fixed by evaluating the inner
products with all Virasoro descendants.}
\begin{eqnarray}
{\cal N}_8 & = & 
h_{ijk} W^i_{-4} W^j_{-2} W^k_{-2} \Omega - 
H_0 \Bigl[ 
\frac{503352}{8072203} 
\, L_{-8} \Omega 
+ \frac{81048}{8072203}\, L_{-6} L_{-2} \Omega \nonumber \\
& & \qquad 
+ \frac{34565}{8072203} \, L_{-5} L_{-3} \Omega 
+ \frac{26403}{16144406} \, L_{-4} L_{-4}\Omega 
+ \frac{110221}{96866436} \, L_{-4} L_{-2} L_{-2} \Omega \nonumber
\\
& & \qquad 
+ \frac{3193}{16144406} \, L_{-3} L_{-3} L_{-2} \Omega 
+ \frac{5210}{24216609} \, L_{-2} L_{-2} L_{-2} L_{-2} \Omega \Bigr]  \ , 
\end{eqnarray}
where $H_0= h_{ijk} h^{ijk}$, which equals in our conventions 
$H_0= 196883\, \tfrac{6929}{6} = \tfrac{1364202307}{6}$.
By the same
token as above (and with somewhat more effort --- in particular, we now also
have to use the null-vector $\N_4$ in order to express the term
$h_{ijk} W^i_{[0]} W^j_{[-2]} W^k_{[-2]}\Omega$ that appears in the
course of this calculation in terms of Virasoro generators) it leads
to the fourth order modular differential equation 
\be\label{Monstd1}
\left[D^4 - \frac{73421}{93780} \, E_4(q) \, D^2 
+ \frac{527029}{562680}\, E_6(q) \, D 
- \frac{1259}{2605} E_4^2(q)
\right] \chi_{M}(q) = 0 \ . 
\ee
This differential equation is actually linearly independent from the other
fourth order modular differential equation of the Monster theory, 
namely the one coming from the null-vector $L_{-2} \N_6$. The latter 
differential equation equals
\be\label{Monstd2}
\left[D^4 - \frac{290}{279} \, E_4(q) \, D^2 
+ \frac{722}{837}\, E_6(q) \, D 
- \frac{8}{31} E_4^2(q)
\right] \chi_{M}(q) = 0 \ ,
\ee
which is in fact simply equal to the $D$-derivative of (\ref{Monstd}). 
Taking the difference of (\ref{Monstd1}) and (\ref{Monstd2})  the Monster
theory therefore also satisfies a modular differential equation of order two,
\be\label{MonstMero}
\left[E_4(q) D^2 
+ \frac{71}{246}\, E_6(q) \, D 
- \frac{36}{41} E_4^2(q)
\right] \chi_{M}(q) = 0 \ .
\ee
[Another way of saying this, is that this is the modular differential equation that 
comes from the nullvector 
\be
{\cal M}_8 = (2\pi i)^{-8}\frac{16151}{2982996}
\Bigl(\frac{24216609}{5210\, H_0}{\cal N}_8
 - L_{-2} \, {\cal N}_6 \Bigr) \ .\Bigr]
\ee
Note that the existence of this second order modular differential equation
is not in conflict with what was said above (or in \cite{Gaberdiel:2007ve}),
since (\ref{MonstMero}) is not holomorphic in the above sense: if we divide 
by $E_4(q)$ to obtain a differential equation whose leading term is $D^2$, the 
coefficient of the term proportional to $D$ is not holomorphic but only meromorphic. 
If  we allow for meromorphic coefficients, every self-dual conformal
field theory obviously also satisfies a first order modular
differential equation (see also \cite{Bantay1}).

\subsubsection{Tensor products of  Monster theories}

Now let us turn to the case of the tensor product of two Monster theories.
(As we shall see  momentarily, the answer for the tensor product of an
arbitrary  
number of Monster  theories can be understood once we have done so for the 
two-fold tensor product.) It is not difficult to show that if (\ref{main2}) was
true, an order $s$ modular differential equation for the tensor
product of the two  
Monster theories would imply that 
\be\label{z2}
\left( L^{(1)}_{-2} + L^{(2)}_{-2}\right)^s \Omega \in O_{[2]} \ .
\ee
Given the arguments of \cite{Gaberdiel:2007ve,Gaiotto:2008jt}   it is
easy to see 
that (\ref{z2}) can only hold for $s\geq 5$. On the other hand, one
finds that the  tensor product of two Monster theories actually
satisfies a fourth 
order differential equation  \cite{Gaiotto:2008jt}, namely
\be \label{MsquaredD4}
\left[D^4 - \frac{175117}{45756} E_4(q)\, D^2 
+ \frac{47539165}{11255976} \, E_6(q) \, D 
- \frac{12838}{52111}\, E_4^2(q) \right] \chi_M^2(q) = 0 \ .
\ee
We now want to explain how to obtain this differential equation
from a null vector in the vacuum Verma module. First we observe 
that the leading term $D^4$ in (\ref{MsquaredD4}) comes from the
vector 
\begin{eqnarray} \label{expandL4}
\left(L_{[-2]}^{(1)} + L_{[-2]}^{(2)} \right)^4 \Omega &  = & 
\Bigl[\left(L_{[-2]}^{(1)}\right)^4+  
4\Bigl(L_{[-2]}^{(1)}\Bigr)^3 L_{[-2]}^{(2)}  \\
& & \qquad  +6\Bigl(L_{[-2]}^{(1)}\Bigr)^2\Bigl(L_{[-2]}^{(2)}\Bigr)^2
+4 L_{[-2]}^{(1)} \Bigl(L_{[-2]}^{(2)}\Bigr)^3
+\Bigl(L_{[-2]}^{(2)}\Bigr)^4 \Bigr]\Omega\ . \nonumber
\end{eqnarray}
In the following we want to show 
how this vector can be expressed, up to terms
of lower conformal weight, in terms of elements in $O_q(V)$.  The terms
in $O_q(V)$ vanish inside any trace, and the terms of lower conformal
weight can be expressed in terms of Virasoro generators, and hence
give rise to the lower coefficients in (\ref{MsquaredD4}).\footnote{Strictly
speaking we also have to guarantee that the resulting terms of lower
conformal weight can be expressed in terms of powers of 
$(L_{[-2]}^{(1)} + L_{[-2]}^{(2)})$, but this can indeed be arranged --- 
this is again a consequence of the fact that there are two independent
null-vectors at level eight.} 

The various terms in (\ref{expandL4}) can now be  rewritten 
as follows. First of all, we observe that every element in $O_q(V)$ is of 
the form
\be
O_q(V) : \quad v + \sum_{n\geq 2} G_n(q) v_n\ , \qquad \hbox{where} \quad
v\in O_{[2]} \ ,
\ee 
and that for any $v\in O_{[2]}$, there is such an element in 
$O_q(V)$.  We call $v$ the `head', and the remaining terms the `tail'. 
Note that the conformal weights of the terms in the tail are always strictly 
smaller than that of $v$. 

Now we can use  the 
null vector $\N_8$ (or $L_{[-2]}\N_6$) to express  $(L_{[-2]}^{(i)})^4\Omega$, 
where $i=1,2$, in terms of a vector in $O_{[2]}$. This can be taken to form
the head of an element in $O_q(V)$, and hence we can rewrite
$(L_{[-2]}^{(i)})^4\Omega$,  up to elements of lower conformal weight 
that come from the tail, as an element of $O_q(V)$. Similarly, we can
reduce $(L_{[-2]}^{(1)})^3 L_{[-2]}^{(2)}\Omega$ by using the 
null-vector 
$\N_6^{(1)}\otimes L_{[-2]}^{(2)}\Omega$, and likewise for the 
term $L_{[-2]}^{(1)} (L_{[-2]}^{(2)})^3\Omega$. The only
difficult term is $(L_{[-2]}^{(1)})^2\, (L_{[-2]}^{(2)})^2\Omega$ for
which this is not possible --- in fact, this is the reason why
(\ref{z2}) with $s=4$ does not hold. We now want to explain 
how this can be circumvented by making use of the null vector $\M_8$. 

As we have seen above, the single Monster theory has a null-vector
at level $8$, $\M_8$, that lies entirely inside $O_{[2]}$, 
$\M_8\in O_{[2]}$. Let us denote by $O_{\M_8}$ its tail, so that 
$\M_8 + O_{\M_8} \cong O_{\M_8} \in O_q(V)$; this is explicitly given
(up to an overall normalisation) as 
\be
O_{\M_8}
= \Bigl[ G_4(q) L_{[-2]}^2 -  \frac{497}{41} G_6(q) L_{[-2]} 
- \frac{26412}{41} G_4(q)^2 \Bigr] \Omega \ ,
\ee
where we have made use of the null vector $\N_4$ at level four
to rewrite the term $W^i_{[-2]}W^i_{[-2]}\Omega$ that appeared in the
course of this calculation in terms of $L_{[-2]}^2\Omega$. 
\smallskip

The same argument also applies to the null vector
$\M_{10}:=L_{[-2]}\M_8$. Up to an overall constant, 
its tail is
\be
O_{\M_{10}}=\Bigl(G_4(q)L_{[-2]}^3 
+ \lambda_1 G_6(q) L_{[-2]}^2 
+ \lambda_2 G_4^2(q) L_{[-2]} 
+  \lambda_3 G_4(q) G_6(q) \Bigr) \Omega \ ,
\ee 
where 
\be
\lambda_1 = \frac{2334255}{1158254} \ ,\qquad 
\lambda_2 = -\frac{451255338}{579127}  \ , \qquad
\lambda_3 = - \frac{10493019690}{579127}\  ,
\ee
and we have used the null-vector relation 
$\widehat{\cal N}_6=0$ with
\begin{eqnarray}
\widehat{\cal N}_6 & = & 
h_{ijk} W^i_{-2} W^j_{-2} W^k_{-2} \Omega
- H_0 \Bigl[\frac{20403}{196883} \, L_{-6} \Omega
+ \frac{5607}{393766} \, L_{-4} L_{-2} \Omega  \\
& & \qquad \qquad\qquad\qquad\qquad\qquad
+ \frac{279}{196883} \, L_{-3} L_{-3} \Omega
+ \frac{8837}{2362596}\, L_{-2} L_{-2} L_{-2} \Omega \Bigr] \ . \nonumber
\end{eqnarray}
Now we can combine these null-vectors to write
\begin{multline} \label{oqholo}
\Bigr[\Bigl(L_{[-2]}^{(1)}\Bigr)^2\Bigl(L_{[-2]}^{(2)}\Bigr)^2  
+\frac{497}{41\, \lambda_1} \Bigl(L_{[-2]}^{(1)}\Bigr)^3 
\Bigl(L_{[-2]}^{(2)}\Bigr) \Bigr] \Omega
+ \hbox{terms of lower conformal weight}
\\
= \frac{1}{G_4(q)} \left\{
\left(L_{[-2]}^{(1)}\right)^2 \Omega^{(1)} \otimes O^{(2)}_{\M_8}
  + \frac{497}{41\, \lambda_1} \, O^{(1)}_{\M_{10}}\otimes 
  \left(L_{[-2]}^{(2)}\right) \Omega^{(2)}
\right\} \in O(q)\ .
\end{multline}
Generically, such an identity will involve coefficients that are not
holomorphic in $q$, since the terms in the bracket on the
right-hand-side will not automatically be divisible by
$G_4(q)$. However, for the specific linear combination that we have
chosen --- \ie for the relative coefficient 
$\tfrac{497}{41 \lambda_1}$ ---  the expression is actually
holomorphic. To see this we observe that the 
coefficients that appear in the bracket are proportional to Eisenstein
series $G_n$ with $n=6,8,10$. Except for $G_6$, these Eisenstein
series are automatically divisible by $G_4$. Thus we only need to
guarantee that the coefficient of $G_6$ vanishes, and this is  
precisely achieved by the above linear combination. 

Translating this analysis back into the language of 
section~3.2, it is now clear that the identity (\ref{nullv})
that corresponds to the fourth order modular differential
equation (\ref{MsquaredD4}) is of the form
\begin{equation}\label{m2o4}
\left( L^{(1)}_{-2} + L^{(2)}_{-2}\right)^4 \Omega + 
\sum_{i=0}^{3} g_i(q) \left( L^{(1)}_{-2} + L^{(2)}_{-2}\right)^i \Omega  
= \frac{1}{G_4(q)} H(q) + \sum_l \hat{h}_l(q) H^l(q) \ ,
\end{equation}
where $H(q)$ is the element of $O_q(V)$ defined by the curly bracket in
(\ref{oqholo}), and the other $\hat{h}_l(q)$ are holomorphic. In this
case the functions $h_j(q)$ of (\ref{nullv})  involve thus one meromorphic (but 
not holomorphic) function, namely $1/G_4(q)$. 

It should be clear from this analysis that such a non-holomorphic
coefficient function $h_j$ in (\ref{nullv}) can only appear if the theory 
has  sufficiently many null-vectors to guarantee that all
non-holomorphic terms on the right-hand-side of (\ref{nullv}) are actually zero. 
(In the above case we had to use, for both theories,  the null-vector 
at level four, the two null-vectors at level six, and the null vector 
at level eight.) For larger conformal  
weight the situation becomes even more constraining since then the
tail will generically also involve Eisenstein series $G_n$ with
$n>14$, none of which are divisible by $G_4$. Thus there will be even
more coefficients that will need to be cancelled! 
\bigskip

Finally, let us comment on the question of how this analysis generalises
to higher tensor powers of the Monster theory. It is clear from the above
analysis that for the $k$-fold tensor product we can always construct a 
modular differential equation of order $k+2$. To see this we expand out 
\be
\left( \sum_{i=1}^{k} L_{-2}^{(i)} \right)^{k+2} \Omega  \ .
\ee
Then each term will either be proportional to $(L_{-2}^{(i)})^3\Omega$ 
for some $i$ --- such terms lie in $O_{[2]}$ by virtue of the null-vector
$\N_6$ --- or  to terms of the form $(L_{-2}^{(i)})^2 (L_{-2}^{(j)})^2 \Omega$
which can be dealt with as explained above. Thus using the above
methods we can construct a modular differential equation at order $k+2$.
On the other hand, this seems to be the minimal order for which such
a differential equation exists \cite{Gaiotto:2008jt}. Thus there do not
seem to be any additional cancellations beyond what is already visible
for the case of the tensor product of two Monster theories. Finally, 
we should stress that the $k$-fold tensor product has a plethora of
low-lying null-vectors: there are at least $k$ linearly independent null-vectors
at level $4$, $2k$ at level $6$, $k$ additional ones at level $8$, {\it etc},
that are relevant for this analysis.

\section{Application to extremal self-dual CFTs}\label{extremal}
\setcounter{equation}{0}

In this final section we want to comment on the implications
of these considerations for the existence of the 
extremal self-dual conformal field theories at 
$c=24k$ that were proposed by Witten \cite{Witten:2007kt}. As was
shown in \cite{Gaberdiel:2007ve}, these theories satisfy a modular
differential equation of degree $s$ where, for large $k$, 
$s\sim \sqrt{k}$. 

As we have shown in section~3 above, every modular differential
equation comes from a null-vector in the vacuum Verma module, see
(\ref{nullv}). Provided that $f_l$ and $h_j$ are holomorphic for 
$|q|<1$, the null-vector relation (\ref{nullv}) implies that (\ref{main2}) 
holds. We now want to show that (\ref{main2})  leads to a 
contradiction for $k\geq 42$. Thus the extremal conformal
field theories can only be consistent for large $k$, provided that
the assumption about the analyticity of $f_l$ and $h_j$ is not
satisfied; we shall comment on this possibility further below.
\smallskip

Suppose then the extremal conformal field theories have a
`null-vector-relation' of the form (\ref{main2}) at conformal weight
$2s$. For $k\geq 42$ this relation is at $L_{[0]}$-weight $2s\leq k$,
and thus arises at a weight where the proposed conformal field theory
only possesses Virasoro descendants of the vacuum.  This then leads to
a contradiction: by the above argument, the right hand side can only
contain Virasoro operators, which we may bring to the standard  
Poincar\'e-Birkhoff-Witt basis. We now claim that no term 
$(L_{[-2]})^s\Omega$ can arise in the process.
Consider first the terms $a_{[-h(a)-1]}b$. Since $b$ can only be a
Virasoro descendant of the vacuum, we can write it as a sum of terms  
\be
L_{[-n_1]}\cdots L_{[-n_N]}\Omega\ ,
\ee
where all $n_l \geq 2$. Since the level of $b$ is $h(b)$, we have
necessarily that $N\leq \lfloor \frac{h(b)}{2} \rfloor$, where
$\lfloor \cdot \rfloor$ denotes the truncated part. Similar
statements also hold for $a$. We now have to the evaluate the
$(-h(a)-1)$-th mode of $a$ and apply it to $b$. The crucial point is that
this mode contains at most as many $L_{[-n]}$ as $a$, see \eg
\cite{Gaberdiel:2000qn}. $a_{[-h(a)-1]}b$ thus has at most 
$\lfloor \frac{h(b)+h(a)}{2} \rfloor 
= \lfloor s - \frac{1}{2}\rfloor = s-1$ $L_{[-n]}$. Since going to the
standard basis only decreases their number, it is clear that we cannot
obtain $(L_{[-2]})^s\Omega$ from this term. 

If we apply the same argument to $d_{[-h(d)+1]}e$, it seems that we
could obtain $s$ Virasoro operators. Note however that $d_{[-h(d)+1]}$
annihilates the vacuum and must therefore contain at least one
$L_{[-n]}$ with $n\leq 1$. Bringing this operator to the right,
commuting through the modes of $e$, we decrease the number of Virasoro
operators at least by one, so that we are again left with 
at most $s-1$ Virasoro generators.

It therefore follows that the right hand side of (\ref{main2}) 
does not contain the term $(L_{[-2]})^s\Omega$. To satisfy the
equality the theory must therefore have a non-trivial null-vector.
At $c>1$, however, we know that the pure 
Virasoro theory does not have any non-trivial null-vectors. 
This then leads to the desired contradiction.

\subsection{A way out?}

This leaves us with the possibility that (\ref{nullv}) does not
imply (\ref{main2}), {\it i.e.}\ that $f_l$ and $h_j$ are not
holomorphic for $|q|<1$. As we have seen in section~3.3,
this can only be the case if the theory has many additional
null-vectors (that guarantee that all coefficients of the meromorphic
functions that would generically appear are actually zero). It 
is certainly conceivable that this can be achieved with only 
null-relations at $h>k$,\footnote{This is not, though, what 
happened in the example of the tensor products of 
the Monster theories: there the null-vectors that are responsible for
this cancellation appear at or below the level suggested by the order
of the differential equation.} and we do not have any hard argument 
against this possibility. There is however a curious observation
that seems to throw some doubt on this scenario.

As we have explained above, the extremal theory at $k=1$, the
Monster theory, has many low-lying null-vectors. This 
property is something one can actually read off from the 
character. To explain this, let us recall that the partition function 
of the Monster theory is 
\be\label{MonsterZ}
Z_M(q) = q^{-1} + 196884\, q 
+ 21493760 \, q^2
+ 864299970 \, q^3
+ \cdots \ .
\ee
We can read off from this formula that there are $N_1=196884$ 
states at level two; these consist of the stress energy tensor $L$,
as well as the fields $W^i$ we have introduced before. 
Now consider the $N_1^2=38\, 763\, 309\, 456$
states 
\be\label{4sta}
L_{-2} L_{-2} \Omega \ , \qquad L_{-2} W^i_{-2} \Omega \ ,  \qquad
W^i_{-2} L_{-2} \Omega\ , \qquad 
W^i_{-2} W^j_{-2} \Omega \ .
\ee
These states appear at level four. On the other hand, we know
from the partition function (\ref{MonsterZ}) that the total number
of states at conformal weight four (above the vacuum) is 
\be
M_1 = 864\, 299\, 970 \ll  38\, 763\, 309\, 456  = N_1^2\ .
\ee
Thus it follows from this simple counting argument that
there must be many `null'-relations among the states (\ref{4sta}); 
one of them is for example the null-vector relation (\ref{n4}). 
\smallskip

One may ask how this counting argument works for the other
extremal self-dual theories. For general $k$ we define 
$N_k$ and $M_k$ by 
\be
Z_k(q) = q^{-k} + \cdots + N_k \, q + \cdots + M_k \, q^{k+2} \ ,
\ee
where $Z_k(q)$ is the extremal partition function. By the same
token as above, the theory will have many null-vectors if 
$M_k - N_k^2<0$. For the first few values of $k$ we find the 
following numbers:
\begin{table}[hbt]
\begin{center}
\begin{tabular}{|c|c|c|c|} \hline
$k$ & $N_k$ & $M_k$ & $M_k - N_k^2$ \\ \hline
&&& \\[-12pt]
k=1 & 196884 & 864299970 & -37899009486 \\ \hline
k=2 & 42987520 & 802981794805760 & -1044945080944640 \\ \hline
k=3 & 2593096794 & 378428749730548169825 & 
371704598747495091389 \\ \hline
k=4 & 81026609428 & 141229814494885904705260482 & 
141223249183450507046773298 \\ \hline
\end{tabular}
\end{center}
\caption{The coefficients $N_k$ and $M_k$ for the 
extremal self-dual theories at $c=24 k$.}
\end{table}

We have checked these numbers for up to $k=150$, and the pattern seems
to continue --- in fact it appears that 
$N_k^2\leq d_1 \, e^{-d_2 k} \,M_k$ for some 
constants $d_1$ and $d_2$. 
Thus this counting argument explains why the Monster theory
has many low-lying null-vectors. It also predicts that the same
is true for the theory with $k=2$, but at least from this point 
of view, there are no indications that the theories with $k\geq 3$
should have many low-lying null-vectors. We regard this as
evidence against the possibility that the extremal theories avoid
the above contradiction.

\section{Conclusions}

In this paper we have shown that every modular differential equation
of a rational conformal field theory comes from a non-trivial null
vector in the Verma module --- see (\ref{nullv}). Generically, the
functions $f_l$ and $h_j$ that appear in this identity are analytic in
$|q|<1$, and then (\ref{nullv}) implies that there is a 
relation of the form (\ref{main2}). At least for the extremal
self-dual theories at $c=24k$ this relation is a non-trivial null
relation. This then implies, following the arguments of
\cite{Gaberdiel:2007ve}, that these theories are inconsistent for
$k\geq 42$. 

This analysis is however not completely conclusive since it {\em is}
possible that the functions $f_l$ and $h_j$ appearing in (\ref{nullv})
are non-holomorphic --- indeed, this is what happens for the
example of Gaiotto \cite{Gaiotto:2008jt} concerning tensor products of
the Monster theory (see (\ref{m2o4})). However, this then requires that the
non-holomorphic terms that appear on the right-hand side of
(\ref{nullv}) must actually vanish, thus indicating that there are
many other null vector relations (albeit none of the form
(\ref{main2})). This is indeed what happens for the case of the tensor
product of the Monster theories. 

Finally, we have seen from the analysis of the partition functions,
that the theories at $k=1,2$ must have many non-trivial null-vector
relations, but that there are no indications (from this point of view)
that this should be the case for $k\geq 3$. Taken together we regard
this as suggestive evidence for the assertion that the extremal
self-dual theories at $c=24k$ are inconsistent for $k\geq 42$. 
\smallskip

The above analysis concerns the extremal bosonic theories at
$c=24k$. It is also interesting to study the supersymmetric
generalisations of this set-up; the case with $N=1$ superconformal
symmetry was already analysed in \cite{Witten:2007kt}, and we have
recently (in collaboration with others) studied the case with $N=2$
superconformal symmetry \cite{GGKMO}. In this case the constraints of
modular invariance are somewhat stronger since one can not only impose
modular invariance of the partition function, but also of the elliptic
genus. In fact, using these constraints one can show that the $N=2$
extremal self-dual theories are inconsistent, except for a few
small sporadic values of the central charge \cite{GGKMO}. One can
also study their modular differential equation, in analogy with what
was done in \cite{Gaberdiel:2007ve}; this will be reported elsewhere.

\section*{Acknowledgements}

This research has been partially supported by the Swiss National
Science Foundation and the Marie Curie network `Constituents,
Fundamental Forces and Symmetries of the Universe'
(MRTN-CT-2004-005104). We are indebted to Terry Gannon for many useful
discussions about these issues.

\appendix
\renewcommand{\theequation}{\Alph{section}.\arabic{equation}}
 
\section{Vertex operator algebras and Zhu's algebra}\label{app:zhu}
\setcounter{equation}{0}

The vacuum representation of a (chiral) conformal field theory
describes a meromorphic conformal field theory \cite{pg}. In 
mathematics, this structure is usually called a vertex operator algebra
(see for example \cite{FLM,Kac} for a more detailed introduction). 
A vertex operator algebra is a vector space 
$V = \bigoplus_{n=0}^\infty V_n$ of states, graded by the conformal
weight. An element $a$ in $V$ of grade $h_a$ defines a linear map on $V$ 
via
\be\label{modex}
a \mapsto  V(a,z) = \sum_{n\in \mathbb{Z}} 
a_n\, z^{-n-h_a}\qquad (a_n \in {\rm End}\ V)\ .
\ee
In this paper we follow the usual physicists' convention for the 
numberings of the modes; this differs by a shift by $h_a-1$
from the standard 
mathematical convention that is also, for example, used in
\cite{Zhu}. 
We also use sometimes (as in \cite{Zhu}) the symbol
\be
o(a) = a_0 \ . 
\ee
Every meromorphic conformal field theory contains an
energy-momentum tensor $L$  with modes
\be
V(L,z) = \sum_n L_n \, z^{-n-2}\ .
\ee
The modes $L_n$ satisfy the Virasoro algebra.

Since much of our analysis is concerned with torus amplitudes
it will be convenient to work with the modes that naturally appear
on the torus; they can be obtained via a conformal transformation
from the modes on the sphere. More specifically, if $a$ is primary, 
we define (see section 4.2 of \cite{Zhu}) 
\be\label{A4}
V[a,z] = e^{2\pi i z h_a}\, V(a,e^{2\pi i z}-1) 
= \sum_n a_{[n]}\, z^{-n-h_a} \ . 
\ee
The explicit relation is then 
\be\label{a[m]}
a_{[m]} = (2\pi i)^{-m-h_a}\sum_{j\geq m} c(h_a,j+h_a-1,m+h_a-1) \,a_j \ , 
\ee
where 
\be
(\log(1+z))^m (1+z)^{h_a-1} = \sum_{j\geq m} c(h_a,j,m) \, z^j \ . 
\ee
This defines a new vertex operator algebra with a new Virasoro
tensor  whose modes $L_{[n]}$ are given by
\be\label{A14}
L_{[n]}= (2\pi i)^{-n}\sum_{j\geq n+1} c(2,j,n+1) L_{j-1} -
(2\pi i)^2\frac{c}{24}\delta_{n,-2} \ .
\ee
The appearance of the correction term for $n=-2$ is due to the fact that
$L$ is only quasiprimary, rather than primary. Furthermore, relative to (\ref{a[m]}) we
have rescaled the modes by a factor or $(2\pi i)^2$ --- this is necessary in order for
the new modes to satisfy the Virasoro algebra (with the same central charge).

\subsection{Zhu's algebra}

One of the key results of Zhu \cite{Zhu} is his characterisation of 
the highest weight representations of a vertex operator algebra 
in terms of representations of an associative algebra $A(V)$, 
usually now called Zhu's algebra. This algebra is defined as the
quotient space of $V$ by the subspace $O_{[1,1]}$, where 
$O_{[1,1]}$ is  spanned by elements of the form 
\be\label{o11}
\oint dz\, \left( V(a,z)\, \frac{(z+1)^{h_a}}{z^2} b \right)\ .
\ee
This definition is motivated by the observation (see for example
\cite{Gaberdiel:1998fs,Gaberdiel:1999mc} for a more detailed exposition) 
that 
\be
\Bigl\langle \phi_1 \Bigl| \phi_2(1) \oint dz\, \left( V(a,z)\, 
\frac{(z+1)^{h_a}}{z^2} b \Bigr) \right\rangle= 0 
\ee
provided only that $\phi_1$ and $\phi_2$ are highest weight
states, {\it i.e.}\ are annihilated by all $a_n$ with $n>0$. 
Thus any combination of two highest weight states defines an
element in the dual space of $A(V)$. Zhu showed that 
also the converse is true; more specifically he proved that 
$A(V)$ carries the structure of an associative algebra
with product 
\be\label{zhupro}
a*b = \oint dz\, \left( V(a,z)\frac{(1+z)^{h_a}}{z} b \right)\ ,
\ee
and that the representations of this associative algebra are in
one-to-one correspondence with the highest weight representations
of the vertex operator algebra. The product structure (\ref{zhupro}) 
describes the multiplication of the zero modes on highest weight states;
in particular, if $\psi$ is a highest weight state, then 
\be
o(a) o(b) \, \psi = o(a*b) \, \psi \ .
\ee
For future reference we also note that in Zhu's algebra (see
\cite{Zhu}, p.296) 
\be\label{commutator}
a*b-b*a = \frac{1}{2\pi i} a_{[-h_a+1]} \, b\ .
\ee
Finally, if $V$ is a rational vertex operator algebra, $A(V)$ is
a semisimple algebra.

\subsection{The $C_2$ space}

The states of the form (\ref{o11}) are not homogeneous with respect to
the $L_0$ grading, even if $a$ and $b$ are. The `leading term', \ie
the term with the highest conformal weight is the term of the form
$a_{-h_a-1}b$. Let us denote the subspace that is generated by states
of this form by $O_{[2]}$. (We are using here the same conventions as
in \cite{Gaberdiel:2000qn}.) A vertex operator algebra is said to
satisfy the $C_2$ criterion if the quotient space $A_{[2]}=V/O_{[2]}$
is finite dimensional. It is easy to see (and proven in \cite{Zhu})
that the $C_2$ condition implies that Zhu's algebra is finite
dimensional. In fact, the dimension of $A_{[2]}$ provides an upper
bound on the dimension of Zhu's algebra. Actually, in many cases
these two dimensions agree, but this is not always the case: in
particular, the dimension of the $C_2$ space is always at least two 
\cite{Gaberdiel:2007ve}, while Zhu's algebra is for example
one-dimensional for self-dual theories. 

Similarly the $C_2$ condition also implies that 
$A_q=V(q)/O_q(V)$ has finite dimension as a
$\mathbb{C}[G_4(q),G_6(q)]$-module \cite{Zhu}. To
see this, we prove the following lemma that
will be useful for the detailed argument in appendix~D.

\noindent {\bf Lemma.} Let $\psi_i,\ i=1,\ldots N$ be a basis of $A_{[2]}$.
Each $v \in V$ can then be written as 
\be \label{Lemma1}
v = \sum_{i=1}^N \lambda(q)_i \psi_i + \sum_\kappa \mu_\kappa
H_\kappa(q) \ , \qquad H_\kappa(q) \in O_q(V)\ ,
\ee
where the sum over $\kappa$ is finite and $\lambda_i(q)$ is a
polynomial in $\mathbb{C}[G_4(q),G_6(q)]$. 

\noindent {\bf Proof}. We 
note that by construction any $v\in V$ can be written as  
\be
v = \sum_{i=1}^{N} \tilde \lambda_i \psi_i + \sum_l r_l\ , \qquad r_l \in O_{[2]}\ ,
\ee
where the sum over $l$ is finite. Since each 
$r_l$ is in $O_{[2]}$, it can serve as the 'head' of an element
$H_l(q) \in O_q(V)$, \ie we can write it as
\be
r_l = H_l(q) + \sum_{k\geq 2} G_{2k}(q)\, \hat r_{l,k}\ ,
\ee
where the states $\hat r_{l,k}$ appearing in the 'tail' have a conformal
weight 
which is lower by at least 4. We can thus apply the same procedure
again and write $\hat r_{l,k}$ as a sum of $\psi_i$ and elements of
$O_{[2]}$. Since the conformal weight decreases in each step, this
algorithm terminates after a finite number of steps. This shows
the Lemma.
\smallskip

Finally we note that since the vertex operator algebras defined by $a_n$ and 
$a_{[n]}$ are isomorphic, the
$C_2$ condition (formulated for either $a_{-h_a-1} b$ or
$a_{[-h_a-1]}b$) implies that the 
${\mathbb C}[G_4(q),G_6(q)]$-ideal of $O_q(V)$ in
$V[G_4(q),G_6(q)]$ has finite codimension. From this it follows that 
there is a relation of the type (\ref{oqstate}).

\section{Torus recursion relations}\label{app:torus}
\setcounter{equation}{0}

In this appendix we briefly sketch the derivation of the recursion
relation (\ref{rec1}); for the detailed argument see \cite{Zhu}.
Let us introduce the notation
\be
F_\H\Bigl((a^1,z_1), \ldots , (a^n,z_n);q\Bigr) 
= z_1^{h_1} \ldots z_n^{h_n}
\Tr_\H \Bigl( V(a^1,z_1) \cdots V(a^n,z_n)\, q^{L_0} \Bigr)\ .
\ee
The derivation of (\ref{rec1}) consists of several steps. We first 
need the following proposition:
\begin{eqnarray}
& F_\H & \hspace*{-0.2cm} 
\Bigl( (a^1,z_1),(a,w),(a^2,z_2),\ldots,(a^n,z_n);q \Bigr)
 \nonumber \\
& = & \hspace*{-0.2cm}
 z_1^{h_1}\ldots z_n^{h_n} \Tr_\H\Bigl( o(a) V(a^1,z_1)\ldots
  V(a^n,z_n)\, q^{L_0} \Bigr) \nonumber \\
& & \hspace*{-0.2cm}
+ \sum_{m\in\Nop_0}   {\cal P}_{m+1}
\left(\frac{z_1}{w},q\right)\, 
\times  F_\H\Bigl( (a_{[m-h_a+1]}a^1,z_1),(a^2,z_2),\ldots,(a^n,z_n);q \Bigr)
 \label{prop1}  \\
& & \hspace*{-0.2cm}
+ \sum_{j=2}^{n}\, \sum_{m\in\Nop_0} \, 
{\cal P}_{m+1}\left(\frac{z_j}{w},q\right) \times 
F_\H\Bigl( (a^1,z_1),(a^2,z_2),\ldots,
(a_{[m-h_a+1]}a^j,z_j),\ldots , (a^n,z_n);q \Bigr) \ .
\nonumber
\end{eqnarray}
Note that there is actually no difference between the terms in the third
line and the fourth line --- we have only distinguished between them
to clarify the derivation below. In fact, it is easy to show that 
$F_\H$ is actually independent of the order in which the $(a^j,z^j)$
(or $(a,w)$) appear, as must be the case.
\smallskip

\noindent {\it Sketch of proof:} The proof is in principle simple:
expand out $V(a,w)$ in modes as in (\ref{modex}). Commute the zero
mode $o(a)$ to the left to get the
second line in (\ref{prop1}); the commutator will eventually be
absorbed into the  
${\cal  P}_1(\frac{z_1}{w},q)$ of the third line, using 
(\ref{Pperiod}). For the other terms in the mode expansion of $V(a,w)$  
we commute each mode $a_k$ through the other fields, using 
\be
[a_k,V(a^j,z_j) ] = \sum_{m\in\Nop_0} \left(
\begin{array}{cc}
h_a-1+k \\ m \end{array} \right) \,
V(a_{m-h_a+1} a^j,z_j) \, z_j^{h_a-1+k-m} \ . 
\ee
As $a_k$ is taken past $q^{L_0}$, we pick up 
\be
a_k \, q^{L_0} =  q^{k} \, 
q^{L_0}  a_k \ .
\ee
Thus when $a_k$ comes back to its original position,
it is multiplied by $q^k$. We can therefore solve for the original
expression to get
\begin{eqnarray}
&& \Tr_\H\Bigl(V(a^1,z_1)\, a_k \cdots
  V(a^n,z_n)\, q^{L_0} \Bigr)  \nonumber \\
&&\qquad = \frac{1}{1-q^k} \sum_{j=2}^n\sum_{l\in \mathbb{N}_0}
\binom{h_a-1+k}{l} z_j^{h_a-1+k-l}  \\
&&\qquad \qquad \qquad \times \Tr_\H\Bigl( V(a^1,z_1)\cdots
  V(a_{l-h_a+1}a^j,z_j) \cdots V(a^n,z_n) \, q^{L_0} \Bigr) \nonumber \\ 
&&\qquad \quad + \frac{q^k}{1-q^k} \sum_{l\in \mathbb{N}_0}
\binom{h_a-1+k}{l} z_1^{h_a-1+k-l} 
\Tr_\H\Bigl( V(a_{l-h_a+1} a^1,z_1)\cdots
  V(a^n,z_n) \, q^{L_0} \Bigr) \ .\nonumber
\end{eqnarray}
We can then plug this into the original expansion and use the identity
\begin{multline}
\sum_{l\in\mathbb{N}_0} \sum_{k=1}^\infty \left(\binom{h_a-1+k}{l}
  \frac{1}{1-q^k} x^k + \binom{h_a-1-k}{l}
  \frac{1}{1-q^{-k}} x^{-k}\right) a_{l-h_a+1}a^j\\
= \sum_{m\in \mathbb{N}_0} {\cal P}_{m+1}(x,q)\, a_{[m-h_a+1]}a^j\ ,
\end{multline}
where ${\cal P}_{m+1}(x,q)$ is the Weierstrass function, see appendix~C. 
For the terms with $j\neq 1$, $x=z_j/w$, so that we obtain directly
the last line of (\ref{prop1}). For $j=1$, $x=qz_1/w$, and we apply
(\ref{Pperiod}) to get the third line. Note that for $m=0$ the shift
by $2\pi i$ is exactly compensated by the commutator term that comes
from the second line. $\square$
\medskip

We will now use (\ref{prop1}) to calculate the action of $a_{[-h_a]}$
on one of the inserted operators. We claim that
\begin{eqnarray}
& & F_\H\Bigl( (a_{[-h_a]}a^1,z_1),(a^2,z_2),\ldots,(a^n,z_n);q \Bigr)
\nonumber \\
&& \quad = z_1^{h_1}\ldots z_n^{h_n} 
\Tr_\H \Bigl(o(a)V(a^1,z_1)\cdots V(a^n,z_n) \,
q^{L_0} \Bigr)\nonumber \\
&& \qquad \quad -\pi i \,
F_\H\Bigl((a_{[-h_a+1]}a^1,z_1),(a^2,z_2),\ldots, (a^n,z_n);q\Bigr)
\nonumber \\
& & \qquad \quad + \sum_{k=1}^{\infty} G_{2k}(q) \, 
F_\H\Bigl( (a_{[2k-h_a]}a^1,z_1),(a^2,z_2),\ldots,(a^n,z_n);q\Bigr)
\label{prop2} \\
& & \qquad \quad + \sum_{j=2}^{n}\sum_{m\in\Nop_0}
{\cal P}_{m+1}\left(\frac{z_j}{z_1},q\right)  F_\H\Bigl(
(a^1,z_1),\ldots,(a_{[m-h_a+1]}a^j,z_j),\ldots , 
(a^n,z_n);q \Bigr) \ .\nonumber 
\end{eqnarray}
\medskip

\noindent {\it Proof:}
We can write the first line of (\ref{prop2}) as
\be
\int_C 
w^{-1}\left(\log\left(\frac{w}{z_1}\right)\right)^{-1}
F_\H\Bigl((a,w),(a^1,z_1),\ldots,(a^n,z_n);q\Bigr) dw\ \label{logcontour}
.
\ee
This can be seen by rewriting $a_{[-h_a]}$ in terms of the original
modes, using 
\be
V(a_l\, a^1,z_1) = \oint_{z_1} dw \, (w-z_1)^{h_a + l -1}\, V(a,w)\, V(a^1,z_1) \ 
\ee
and by the definition of the $c(h_a,j,m)$, 
\be
\sum_{j\geq-1}c(h_a,j,-1)\,(w-z_1)^j z_1^{h_a-1-j}w^{-h_a}=
w^{-1}\left(\log\left(\frac{w}{z_1}\right)\right)^{-1}\ .
\ee
We then use (\ref{prop1}) to evaluate $F_\H$. 
From (\ref{logcontour}) we see that in the terms that are regular 
in $w=z_1$,
we simply need to replace $w$ by $z_1$. To evaluate the third line of
(\ref{prop1}) we substitute $z_1= \exp(2\pi i
z_1'), w = \exp(2\pi i w')$, which shows that we
obtain the constant term in the $w'$ expansion of ${\cal
  P}_{m+1}(e^{2\pi i w'})$ , which can be read off directly from
(\ref{Pexpansion}). $\square$

\medskip
\noindent To get (\ref{rec1}), we specialise (\ref{prop2}) to the case
$n=1$. Furthermore we use that (see \cite{Zhu})
\be\label{commutatora}
[o(a),V(z^{L_0}\, b,z)] = (2\pi i )\, V(z^{L_0}\, a_{[-h_a+1]}b,z) \ ,
\ee
implying that $F_\H((a_{[-h_a+1]}b,z);q)=0$. If we consider 
the terms of (\ref{prop2}) of power $z^0$, we thus obtain
\be
{}\Tr_\H \Bigl(o(a_{[-h_a]}b)\, q^{L_0} \Bigr) = 
\Tr_\H \Bigl(o(a) \, o(b) \, q^{L_0} \Bigr)
+ \sum_{k=1}^\infty G_{2k}(q,y) 
    \Tr_\H \Bigl( o(a_{[2k-h_a]}b) \, q^{L_0} \Bigr)\ .
\ee

\subsection{Differential operators} \label{app:diffops}

For the determination of the modular differential equation, one of the
key steps is the calculation of the differential operators $P_s(D)$,
see (\ref{dop}). In the following, we give explicit formulae for them
for the first few values of $s$
\begin{eqnarray}
P_1(D) &=& (2\pi i)^2 D \\
P_2(D) &=& (2\pi i)^4 D^2 + \frac{c}{2}\, G_4(q) \\
P_3(D) &=& (2\pi i)^6 D^3 + \left(8+\frac{3c}{2}
\right) G_4(q) \, (2\pi i)^2 D + 10c\, G_6(q) \\
P_4(D) &=& (2\pi i)^8 D^4 + (32+3c)\, G_4(q) \, (2\pi i)^4
D^2 + (160+40c)\, G_6(q)\, (2\pi i)^2 D \nonumber \\
& &+ \left(108 c + \frac{3}{4}c^2\right)\, G_4(q)^2  \ . 
\end{eqnarray}
Here $c$ is the central charge of the corresponding conformal field
theory.

\section{Weierstrass functions and Eisenstein series}
\setcounter{equation}{0} \label{app:Eisenstein}

Let us define the function
\be
{\cal P}_k(q_z,q) = \frac{(2\pi i)^k}{(k-1)!} \sum_{n=1}^\infty \left(
  \frac{n^{k-1}q_z^n}{1-q^n} + \frac{(-1)^k n^{k-1}
    q_z^{-n}q^n}{1-q^n} \right) \ ,
\ee
which converges for $|q|<|q_z|<1$. Since $q_z\frac{d}{dq_z}{\cal P}_k(q_z,q) =
\frac{k}{2\pi i} {\cal P}_{k+1}(q_z,q)$, we will concentrate on ${\cal
  P}_1(q_z,q)$.  
In what follows, we shall be interested in the
behaviour around $q_z=1$. ${\cal P}^Q_1(q_z,q,y)$ has a simple pole at
$q_z=1$, but we can find a meromorphic continuation on
$|q|<|q_z|<|q|^{-1}$ by rewriting
\be
{\cal P}_1(q_z,q) = \frac{2\pi i}{1-q_z} - 2\pi i +  2\pi i
\sum_{n=1}^\infty \left( \frac{q_z^n q^n }{1-q^n} -
  \frac{q_z^{-n}q^n }{1-q^n}\right) \ .
\ee
A straightforward calculation then shows the
identity
\be
{\cal P}_1(qq_z,q)={\cal P}_1(q_z,q)+2\pi i\ . \label{Pperiod}
\ee
Introducing the new variable $z$ by $q_z = e^{2\pi i z}$, we want to
calculate the Laurent expansion in $z$ around 0. The crucial point is
that the coefficients of 
this Laurent expansion are essentially the Eisenstein series
$G_{2k}(q)$ that will eventually appear in (\ref{rec1}). In fact,
expanding $q^z$ in $z$ and 
using the definition of the Bernoulli numbers,
\be
\frac{x}{e^x-1} = \sum_{n=0}^\infty \frac{B_n}{n!}x^n\ ,
\ee
along with the identity
\be
B_{2n}=\frac{(-1)^{n-1}2(2n)!}{(2\pi)^{2n}} \zeta(2n)\ ,
\ee
we obtain 
\be
{\cal P}_1(q_z,q) =  -\frac{1}{z} -\pi i +
\sum_{k=1}^\infty G_{2k}(q) z^{2k-1}\ , \label{Pexpansion}
\ee
where the Eisenstein series
are defined by
\be
G_{2k}(q) = 2\zeta(2k)+ \frac{2(2\pi i)^{2k}}{(2k-1)!}
\sum_{n=1}^\infty \frac{n^{2k-1} q^n 
    }{1-q^n} \ .
\ee
The Laurent expansions of the higher ${\cal P}_k(q_z,q)$ functions can
be directly obtained by 
\be
\partial_z {\cal P}_k(q_z,q) = k{\cal P}_{k+1}(q_z,q)\ .
\ee

\subsection{The Eisenstein series}

The Eisenstein series $G_{2k}(\tau)$ can also be alternatively defined
by 
\ba
G_{2k}(\tau)&=& \sum_{(m,n)\neq(0,0)} \frac{1}{(m\tau +n)^{2k}} \qquad
  k\geq 2\ ,\\
G_{2}(\tau) &=& \frac{\pi^2}{3} + \sum_{m\in\mathbb{Z}-\{0 \}}
\sum_{n\in\mathbb{Z}} \frac{1}{(m\tau+n)^2}\ . 
\ea
For $k\geq 2$, $G_{2k}(\tau)$ is a modular form of weight $2k$, \ie
\be
G_{2k}\left(\frac{a\tau+b}{c\tau+d}\right) 
= (c\tau +d)^{2k}G_{2k}(\tau)\ ,
\ee
whereas $G_2(\tau)$ transforms with a modular anomaly
\be
G_{2}\left(\frac{a\tau+b}{c\tau+d}\right) 
= (c\tau +d)^{2}G_{2}(\tau) - 2\pi i c(c\tau+d) \ .
\ee
We can use $G_2$ to define a modular covariant derivative: If $f(q)$
is a modular form of weight $s$, then $D_s f(q)$ is a modular form of
weight $s+2$, where
\be
D_s = q\frac{d}{dq} - \frac{s}{4\pi^2}G_2(q)\ . \label{covD}
\ee
The space of modular covariant functions is given by the ring
$\mathbb{C}[G_4(q), G_6(q)]$ that is freely generated by $G_4(q)$ and
$G_6(q)$. In particular, all higher $G_{2k}(q)$ can be written as
polynomials in $G_4(q), G_6(q)$.

It is also sometimes convenient to work with a different normalisation
for the Eisenstein series, so that the constant term is $1$; the
corresponding series will be noted by $E_n(q)$. For the first
few values of $n$, they are explicitly given as 
\begin{eqnarray}
E_2(q) & = & 1-24\, q-72\,  q^2-96\,  q^3-168\,  q^4
-144\,  q^5-288\, q^6 - \cdots
\ , \nonumber \\
E_4(q) & = & 1 + 240 \, q + 2160\,  q^2 + 6720\,  q^3 
+ 17520\,  q^4 + 30240\,  q^5 + 60480\,  q^6 + \cdots \ , 
\nonumber \\
E_6(q) & = & 1 - 504\,  q - 16632\,  q^2 - 122976\,  q^3 
- 532728\,  q^4 - 1575504\,  q^5 - 4058208  q^6 - \cdots .  
\nonumber 
\end{eqnarray}
The relation between the $G_n(q)$ and $E_n(q)$ is simply
$G_n(q) = 2 \zeta(n) \, E_n(q)$;  for the first few values of $n$, we
have explicitly
\be
G_2(q)  =  - \frac{(2\pi i)^{2}}{12}\, E_2(q) \ , \quad
G_4(q) = \frac{(2\pi i)^{4}}{720}\, E_4(q)\ , \quad
G_6(q) = -\frac{(2\pi i)^{6}}{30240}\, E_6(q)\ .
\ee
Finally, we mention that we have the identities
\be
D E_4 = - \frac{1}{3} E_6 \ , \qquad
D E_6 = - \frac{1}{2} E_4^2 \ , \qquad
D E_4^2 = - \frac{2}{3} E_4 E_6 \ .
\ee

\section{Radius of convergence}
\setcounter{equation}{0} \label{app:convergence}

In this appendix we want to explain the details of the calculation leading up
to (\ref{nullv}). We shall also show that if the theory is $C_2$-finite then the
functions $f_l(q)$ and $h_j(q)$ defined in (\ref{ndef1}) have a
non-vanishing radius of convergence.  We begin by deriving some
relations that will be important for the argument in section~D.2. 

\subsection{$A_{[2]}$ relations}

In the following we shall assume that the theory is 
$C_2$ finite.\footnote{If we do not assume $C_2$ finiteness,
the argument can be done essentially the same way, the only
difference being that we cannot show that only finitely many
correction terms appear. The resulting coefficient functions are 
then only formal power series in $q$.}
 This means that the space $A_{[2]} =V/O_{[2]}$ is 
finite-dimensional, say of dimension $N$. We denote 
the irreducible representations of the theory by 
$M_j, j = 1,\ldots,N'$, where $N' \leq N$. Given the close
relation between the $A_{[2]}$ space and Zhu's algebra
(see appendix~A.2) we can then choose a basis 
$\psi_i,\ i=1,\ldots, N$ for $A_{[2]}$ such that
\be\label{bascho}
{}\Tr_{M_j} \psi_i = \delta_{i,j}\ , \quad i=1,\ldots,N'\ , \qquad
\Tr_{M_j} \psi_i = 0\ , \quad \forall j\ ,\ i=N'+1,\ldots, N\ . 
\ee
For the analysis of section~3.2 it is important to obtain
good recursive relations for the vectors that vanish in 
all traces, \ie the vectors $\psi_i$ with $i=N'+1,\ldots N$. 
In a first step we claim that we can write
\be \label{psidecomp}
\psi_i = \sum_l d^{l,i}_{[-h(d^{l,i})+1]} e^{l,i} 
+ \sum_{\kappa \in S_i} \alpha_{(i)}^\kappa H_\kappa(0)\ , \qquad
i=N'+1,\ldots N \ ,
\ee
where each $H_\kappa(q)$ is an element in
$O_q(V)$. To prove this we will assume that Zhu's algebra 
$A(V)$ is semisimple. 
The proof proceeds in two steps. First we show that if any vector
$a$ is trivial in all traces, then $a$ must equal a commutator
in Zhu's algebra. This follows for example from a standard theorem of
associative algebras, the Wedderburn structure theorem \cite{Farb}. It
states that every semisimple associative algebra is isomorphic to the 
product of algebras of $n\times n$ matrices over $\mathbb{C}$, 
\be
A(V) \cong \prod_{i=1}^N {\cal M}_{n_i}(\mathbb{C}) \ ,
\label{Wedderburn} 
\ee
where $n_i$ is the dimension of the $i^{\rm th}$ irreducible
representation $M_i$ of $A(V)$. Assume we are given $a\in A(V)$ such
that $\Tr_{M_i} (a) = 0$ for all irreducible representations $M_i$. By 
(\ref{Wedderburn}), $a$ is isomorphic to a blockdiagonal matrix whose
blocks all have vanishing trace. It is then a straightforward exercise
to show that each such matrix can be written as a sum of
commutators. Because of the identity (\ref{commutator}) we thus find
that, up to elements in $O_{[1,1]}$, we have
\be\label{com1}
2\pi i\cdot a =  2\pi i\, \sum_l (d^{l}*e^l-e^l*d^l) 
= \sum_l d^l_{[-h(d^l)+1]}\, e^l \ .  
\ee

The argument so far implies that up to commutator terms (\ref{com1}), 
$\psi_i$ lies in $O_{[1,1]}$, the subspace by which we quotient to
obtain Zhu's algebra $A(V)$. On the other hand, $O_{[1,1]}$ is closely
related to $O_q(V)$: for any state in $O_q(V)$,
\be
H(q) \equiv a_{[-h_a-1]}b + \sum_{k\geq 2} (2k-1) G_{2k}(q) a_{[2k-h_a-1]}b\ ,
\ee
we can formally take the limit $q\rightarrow 0$, \ie we can consider
its constant part only. Then we obtain (see \cite{Zhu}, Lemma 5.3.2)
\be
H(0) = \frac{\pi i}{6} a_{[-h_a+1]}b + 2\pi i\, \oint dz\, 
\left( V(a,z) \frac{(1+z)^{h_a}}{z^2} b \right)\ , \label{qto0}
\ee
\ie up to a commutator term, the limit is in $O_{[1,1]}$. In fact, it
is obvious that every element in $O_{[1,1]}$ can be obtained in this
manner.  Together with (\ref{com1}) this then proves the claim
(\ref{psidecomp}). We should stress that for each $i$, only finitely
many different $d^{l,i}, e^{l,i}$ and $H_\kappa$ appear.

\subsection{Evaluating $K(q)$}

As in (\ref{kdef}) let 
$K(q)=\sum_r g_r(q) v_r$, $g_r(q) \in \mathbb{C}[G_4(q),G_6(q)]$ 
be such that 
\be
\Tr_{M_j} \Bigl( o(K(q)) \, q^{L_0-\frac{c}{24}} \Bigr)=0\ , \qquad \forall
M_j\ .
\ee
Using the Lemma from appendix~A, (\ref{Lemma1}), we can write
\be
K(q) = \sum_{i=1}^N \lambda_i(q) \psi_i + O_q(V) =: K'(q) + O_q(V)\ ,
\ee
so that from (\ref{psidecomp}) 
\be \label{decomp1}
K'(q)= \sum_{i=1}^N \lambda_i(q) \psi_i
= \sum_{i=1}^{N'} \lambda_i(q) \psi_i  +\sum_{i=N'+1}^N
\lambda_i(q) \sum_l d^{l,i}_{[-h(d^{l,i})+1]} e^{l,i}
+  \sum_{\kappa\in S} \alpha^\kappa(q)
H_\kappa(0) \ ,
\ee
where $\alpha^\kappa(q) =  \sum_{i=N'+1}^N \lambda_i(q)\alpha_{(i)}^\kappa$
and $S = \bigcup_{i=N'+1}^N S_i$ with $S_i$ from (\ref{psidecomp}). Now we
define
\be
N_0(q) =  \sum_{i=N'+1}^N
\lambda_i(q) \sum_l d^{l,i}_{[-h(d^{l,i})+1]} e^{l,i} +  \sum_{\kappa \in S}
\alpha^\kappa(q) H_\kappa(q)
\ee
and
\be
\Delta_0(q) = K'(q)-N_0(q) = \sum_{i=1}^{N'}\lambda_i(q)\psi_i +
\sum_{\kappa \in S} \alpha^\kappa(q)\left(H_\kappa(0)-H_\kappa(q)\right)
\ .
\ee
Since both $K'(q)$ and $N_0(q)$ vanish in all traces for all values of
$q$, also $\Delta_0(q)$ vanishes. Because of our choice of basis
(\ref{bascho}), we know that for $q=0$ each 
$\lambda_i(0)$ with $i=1,\ldots,N'$ vanishes. It is thus possible
to define $\tilde \Delta_0(q) := \Delta_0(q)/q$, which is, by construction, still a
power series in $q$.
\smallskip

\noindent Next we rewrite the second part of $\tilde \Delta_0(q)$ 
(leaving out the coefficient $\alpha^\kappa(q)$ for the moment) as
\begin{multline}
q^{-1}(H_\kappa(0)-H_\kappa(q)) = \sum q^{-1}(G_{2k}(0)-G_{2k}(q))\, v^\kappa_k
= \sum_{i=1}^N \lambda^i_\kappa(q) \psi_i + \sum_{\tau\in T} \tilde
\mu(q)^{\ \tau}_\kappa H_\tau(q) \\
= \sum_{i=1}^{N'} \lambda^i_\kappa(q) \psi_i + \sum_{\tau\in T} \tilde
\mu(q)^{\ \tau}_\kappa H_\tau(q) + \sum_{i=N'+1}^N
\lambda^i_\kappa(q) \sum_l d^{l,i}_{[-h(d^{l,i})+1]} 
e^{l,i} + \sum_{\tau\in S} \mu(q)_\kappa^{\ \tau} H_\tau(0) \ ,
\end{multline}
where in the second equality we have applied the Lemma 
from appendix~A, (\ref{Lemma1}), to each $v_k^\kappa$, and $T$
is the finite set of all elements of $O_q(V)$ that appear in this
process. In the last step we have again used the previous recursion step
for the $\psi_i$ with $i=N'+1, \ldots,N$. 
In particular, the set $S$ is the same as before. 
Since $\tilde \Delta_0(q)$ still vanishes in all traces, we can set
$q=0$ to see that the (total) coefficient of each $\psi_i, i=1,\ldots N'$ vanishes. 
We then define
\be
N_1(q) = \sum_\kappa \alpha^\kappa(q)\left(
\sum_\tau \tilde
\mu(q)^{\ \tau}_\kappa H_\tau(q) + \sum_{i=N'+1}^N
\lambda^i_\kappa(q) \sum_l d^{l,i}_{[-h(d^{l,i})+1]} 
e^{l,i} + \sum_{\tau\in S} \mu(q)_\kappa^{\ \tau} H_\tau(q)
\right)
\ee
and $\Delta_1(q)= \tilde \Delta_0(q)-N_1(q)$. It is clear that we can
apply the same reasoning to $\Delta_1(q)$ and all the subsequent
$\Delta_n(q)$. 
It is important to note that the only $H_\kappa(q)$ that appear 
are those with $\kappa\in S$ or $\kappa \in T$.
In total we thus obtain
the (a priori formal) power series
\be
K'(q) = \sum_{n=0}^\infty q^n N_n(q) = \sum_l f_l(q) \,
d^l_{[-h(d^l)+1]}\, e^l  + \sum_{\kappa\in S\cup T} h^\kappa(q) \, H_\kappa(q) 
\ .
\ee
To show that it has a non-vanishing radius of convergence, 
note that for example the last term of $N_n(q)$ is of the form 
\be
q^n \alpha^\kappa(q)(\underbrace{\mu(q) \cdot \mu(q) \cdots
  \mu(q)}_n)_{\kappa}^{\ \tau}  H_\tau(q)\ .
\ee
By construction, $\mu(q)_\kappa^{\ \tau}$ is holomorphic 
for all $\kappa$ and $\tau$, and thus the $\sup_{|q|<1/2} 
|\mu(q)_\kappa^{\ \tau}|$ is finite, so that the norm $D$ of the matrix
$(\mu)_\kappa^{\ \tau} $ is also finite. It thus follows that the
radius of convergence $\rho$ is at least $\min\{\frac{1}{D},1/2\}$.
The other terms in $N_n(q)$ can be dealt with similarly.

This argument therefore shows that the coefficient functions of 
$H_\kappa(q)$ as well as those of the commutator terms 
$d^{l}_{[-h(d^{l})+1]}  e^l$ have finite radius of convergence.

\end{document}